\documentclass[12pt,preprint]{aastex}

\slugcomment{{\it Astrophysical Journal Letters}, Vol. 571, June 1, 2002, 
in press}

\def\lsim{\mathrel{\rlap{\lower 4pt \hbox{\hskip 1pt $\sim$}}\raise 1pt \hbox
        {$<$}}}
\def\gsim{\mathrel{\rlap{\lower 4pt \hbox{\hskip 1pt $\sim$}}\raise 1pt \hbox
        {$>$}}}

\begin{document}

\righthead{Confronting Neutron Star Cooling Theories with
New Observations}

\title{Confronting Neutron Star Cooling Theories with
New Observations}

\author{S. Tsuruta}
\affil{Department of Physics, Montana State University, Bozeman,
MT 59717, USA}
\email{uphst@gemini.oscs.montana.edu}

\author{M.A. Teter}
\affil{Department of Astronomy and Astrophysics, Pennsylvania
State University, University Park, PA 16802, USA}
\email{teter@lonestar.astro.psu.edu}

\author{T. Takatsuka}
\affil{Faculty of Humanities and Social Sciences, Iwate
University, Morioka, Iwate 020-8550, Japan}
\email{takatuka@iwate-u.ac.jp}

\author{T. Tatsumi}
\affil{Department of Physics, Kyoto University, Kyoto 606-8502,
Japan}
\email{tatsumi@ruby.scphys.kyoto-u.ac.jp}

\and

\author{R. Tamagaki}
\affil{Kamitakano Maeda-cho 26-5, Kyoto 606-0097, Japan}
\email{tamagaki@yukawa.kyoto-u.ac.jp}

\begin{abstract}
With the successful launch of {\it Chandra} and {\it XMM/Newton}
X-ray space missions combined with the lower-energy band
observations, we are now in the position where careful comparison
of neutron star cooling theories with observations will make it
possible to distinguish among various competing theories. For
instance, the latest theoretical and observational developments
appear to exclude both nucleon and kaon direct URCA cooling. In
this way we can now have realistic hope for determining various
important properties, such as the composition, degree of
superfluidity, the equation of state and stellar radius. These
developments should help us obtain better insight into the
properties of dense matter.

\end{abstract}

\keywords{Dense matter --- stars: neutron --- X-rays: stars}

\section{Introduction}

The launch of the {\it Einstein} Observatory gave the first hope
for detecting thermal radiation directly from the surface of
neutron stars (NSs).  However, the temperatures obtained by the
{\it Einstein} were only the upper limits (e.g., Nomoto \& Tsuruta
1986). {\it ROSAT} offered the first confirmed detections (not
just upper limits) for such surface thermal radiation from at
least three cooling neutron stars, PSR 0656+14, PSR 0630+18
(Geminga) and PSR 1055-52 (e.g., Becker 1995).  Very recently, the
prospect for measuring the surface temperature of isolated NSs, as
well as obtaining better upper limits, has increased tremendously,
thanks to the superior X-ray data from {\it Chandra} and {\it
XMM/Newton}, as well as the data in lower energy bands from
optical-UV telescopes such as {\it Hubble Space Telescope (HST)}.
Consequently, the number of possible surface temperature
detections has already increased to at least nine, by addition of
RX J0822-4300 (Zavlin, Tr\"umper, \& Pavlov 1999),  the Vela
pulsar (Pavlov et al. 2001), RX J1856-3754 (Zavlin et al. 2002a),
1E 1027.4-5209 (Zavkin, Pavlov, \& Tr\"umper 1998), RX J0002+62
(Kaminker, Yakoblev, \& Gnedin 2002), and RX J0720.4-3125 (Haberl,
et al. 1997).  Most recently Chandra offered an important upper
limit to PSR J0205+6449 in 3C58 (Slane, Helfand and Murray 2002).
It will be a matter of time when more detections, as well as
better upper limits, will become available. These developments
have proved to offer serious `turning points' for the
detectability of thermal radiation directly from NSs, because we
can now seriously compare the observed temperatures with NS
cooling theories (see., e.g., Tsuruta \& Teter 2001a, 2001b,
hereafter TT01a, TT01b).  On the theoretical side, more detailed,
careful investigations have already started.  In this report we
try to demonstrate that distinguishing among various competing NS
cooling mechanisms already started to become possible, by careful
comparison of improved theories with new observations.

\section{Neutron Star Cooling Models}

The first cooling calculations (Tsuruta 1964) showed that NSs can
be visible as X-ray sources for about a million years.  After a
supernova explosion a newly formed NS first cools via various
neutrino emission mechanisms before the surface photon radiation
takes over (e.g., Tsuruta 1979). Among the important factors which
seriously affect the nature of NS cooling are: neutrino emission
processes, superfluidity of the constituent particles,
composition, and mass.  In this paper, for convenience, the
conventional, slower neutrino cooling mechanisms, such as the
modified URCA, plasmon neutrino and bremsstrahlung processes which
were adopted in many of the earlier and subsequent cooling
calculations, will be called `standard cooling'.  On the other
hand, the more exotic extremely fast cooling processes, such as
the direct URCA processes involving nucleons, hyperons, pions,
kaons, and quarks, will be called `nonstandard' processes (see,
e.g., Tsuruta 1998, hereafter T98).

The composition of the core is predominantly neutrons, with a
small fraction of protons and electrons, if the density is
moderate ($< \sim$ 10$^{15}$ gm/cm$^3$).  For higher densities
hyperons, pions, kaons, quarks, etc., may dominate the central
core.  As the star cools after the explosion and the interior
temperature falls below the superfluid critical temperature,
T$^{cr}$, some hadronic components become superfluid. That causes
exponential suppression of both specific heat (and hence the
internal energy) and all neutrino processes involving these
particles.  The net effect is that the star cools more slowly (and
hence the surface temperature and luminosity will be higher)
during the neutrino cooling era. Very recently, the `Cooper pair
neutrino emission' (Flowers, Ruderman, \& Sutherland 1976) was
rediscovered to be also important under certain circumstances
(see, e.g., Yakovlev, Levenfish, \& Shibanov 1999). This process
takes place right after the core temperature falls below T$^{cr}$,
but it soon falls exponentially. The net effect is to enhance, for
some superfluid models, the neutrino emission right after the
superfluidity sets in.

We have calculated NS cooling adopting the most up-to-date
microphysical input and the fully general relativistic, `exact'
evolutionary code (without making isothermal approximations) which
was originally constructed by Nomoto \& Tsuruta (1987) and has
been continuously up-dated. All possible neutrino emission
mechanisms both in the stellar core and crust, including the
Cooper pair neutrino emission, are used.  The results are summarized
in Fig. 1, which shows the cooling curves of representative NSs
with an equation of state (EOS) of medium stiffness, the FP
Model constructed by Friedman \& Pandharipande (1981). The solid
curve refers to the standard cooling of a 1.2 M$_\odot$ star which
consists predominantly of neutrons (with proton fraction of $\sim$
5 \%). For this star the stellar core density, $\rho_c$, is below
the critical density, $\rho_{tr}$, where the transition to the
pion-condensed phase occurs.  We adopt $\rho_{tr}$ = 2.5 $\rho_0$
taken from Umeda et al. 1994 (hereafter U94), where $\rho_0$ = 2.8
x 10$^{14}$ gm/cm$^3$ is the nuclear density.  For this star we
adopt the T72 superfluid model (Takatsuka 1972) which is based on
the most realistic treatment available for the neutron-dominated
matter. We find that the superfluid effect
is minor.  This is because the superfluid gap (which is
proportional to T$^{cr}$) depends on density, and the gap
disappears at the maximum density, $\rho_{max}$, which is less
than $\rho_c$, the central density of the star. On the other hand,
the 1.4 M$_\odot$ star (dashed curve) cools by the pion cooling as
the nonstandard case. This is because for this star the central
density $\rho_c$ exceeds $\rho_{tr}$,
and hence the central core consists of pion condensates.  As the
superfluid model for the pion-condensed phase we adopt a medium
superfluid gap model, called the E1-0.8 Model (see U94), which is
constructed from the microphysical superfluid gap models
calculated for the pion matter (Takatsuka \& Tamagaki 1982,
hereafter TT82). For this superfluid model both enhanced Cooper
pair neutrino emission right below T$^{cr}$ and subsequent
suppression are significant.  As the stellar mass increases from
1.2 M$_\odot$ to 1.4 M$_\odot$ the central density increases, and
the corresponding stars lie between the solid and dashed curves in
Fig. 1. Further details are found in TT01a, TT01b, Teter 2001, and
Teter \& Tsuruta 2002 (hereafter TT02).

\section{Comparing Neutron Star Cooling Models with New Observations}

In Fig. 1 cooling curves are compared with the latest confirmed
observational data.  We may note that the data in Fig. 1 suggest
the existence of at least two classes of sources, hotter stars
(e.g., (1) RX J0822-4300, (3) PSR 0656+14, (5) PSR RXJ 1856-3754
and (6) PSR 1055-52), and cooler stars (e.g., (c) PSR J0205+6449,
(2) the Vela pulsar and (4) Geminga). The hotter sources are
consistent with the solid curve, the standard cooling of a 1.2
M$_\odot$ star. The source (6) is somewhat above the solid curve,
but that is easily explained when heating is included (see, e.g.,
Umeda, Tsuruta, \& Nomoto 1994, hereafter UTN94) and when the age
uncertainty, of a factor of $\sim$ 2, is taken into account. On
the other hand, the cooler stars are consistent with the dashed
curve, the pion cooling of a 1.4 M$_\odot$ star with significant
superfluid suppression.  (The age uncertainty should not affect
this conclusion for younger sources because the slope of the
curves in these younger years is small.)
The conclusion is that the observed data are all most naturally
explained as the effects of stellar mass and superfluidity of the
constituent core particles.

At least for binary pulsars, recent observations offer stringent
constraints on the mass of a NS, to be very close to 1.4 M$_\odot$
(e.g., Brown, Weingartner, \& Wijers 1996). If this evidence
extends to isolated NSs also, then the EOS should be such that the
mass of the star whose central density is very close to the
transition density (where the nonstandard process sets in) should
be very close to 1.4 M$_\odot$.  With the EOS of our current
choice, medium FP Model, that transition takes place between 1.2
M$_\odot$ and 1.4 M$_\odot$.  Then if the deviation from the
`magic number' 1.4 M$_\odot$ should be very small the correct EOS
may have to be somewhat stiffer than medium.  This is because a
stiffer EOS corresponds to somewhat larger mass, for given
transition density, and hence the corresponding mass for the solid
and dashed curves will become, e.g., 1.35 M$_\odot$ and 1.45
M$_\odot$, respectively, instead of 1.2 M$_\odot$ and 1.4
M$_\odot$. In this way comparison of cooling curves with
observation has a potential for determining the EOS and hence
radius, if the mass is fixed. Note that comparison of this kind
already eliminates very soft and very stiff EOS. The reason is
that a very soft EOS will be so compact and the density so high
that the phase transition will take place for stars of too small
mass, e.g., $\sim$ 0.3 M$_\odot$. On the other hand, the central
density of stars with a very stiff EOS will be below the
transition density, meaning that the dashed curve can never exist.

The qualitative behavior of all nonstandard scenarios is similar
if their transition density is the same (see, e.g., UTN94, T98).
However, here we try to demonstrate that it is still possible to
offer comprehensive assessment of at least which options are more
likely while which are less likely. First of all, we note that all
of the nonstandard mechanisms are too fast to be consistent with
any observed detection data, even with heating (see, e.g., UTN94,
T98). That means {\it significant suppression of neutrino
emissivity due to superfluidity is required}. However, recently
Takatsuka \& Tamagaki (1997)(hereafter TT97) showed, through
careful microphysical calculations, that for neutron matter with
high proton concentration which permits the nucleon direct URCA,
the superfluid critical temperature T$^{cr}$ should be too low,
$\sim$ several x 10$^7$ K, both for neutrons and protons. On the
other hand, the observed NSs, which are to be compared with
cooling curves, are all hotter (the core temperature being
typically $\sim$ 10$^8$ K to several times 10$^8$ K). That means
{\it the core particles are not yet in the superfluid state} in
these NSs. The implication is that {\it these sources would be too
cold if nucleon direct URCA were in operation.}  The same argument
applies to the kaon cooling also (Takatsuka \& Tamagaki 1995,
hereafter TT95).

Recently, Kaminker, Yakovlev, \& Gnedin 2002 (hereafter KYG02)
carried out cooling calculations similar to ours by adopting the
nucleon direct URCA as the nonstandard option.  However, as just
pointed out, {\it the observed NSs adopted for their comparison
would be yet to be in the superfluid state}, and hence these stars
would be too cold if the nucleon direct URCA cooling were in
operation. The calculations by KYG02 led to two important
conclusions concerning the core superfluids: (i) proton superfluid
critical temperature, T$^{cr}_p$, should be relatively high,
$\sim$ 7 x 10$^9$ K, if the nucleon direct URCA is to be in
operation in the observed cooler NSs; while (ii) the neutron
superfluid critical temperature, T$^{cr}_n$, should be less than
10$^8$ K, for the same superfluid model to be consistent with the
data of hotter NSs.  The conclusion (ii) agrees with realistic
microphysical calculations of the neutron matter with high proton
concentration which allows the nucleon direct URCA (TT97), but the
conclusion (i) contradicts with these microphysical results. We
may note that KYG02 did not actually carry out microphysical
calculations, but only assumed that the $^1S_0$ superfluid gap of
neutrons and protons are similar and also that the proton
effective mass is 0.7 m$_p$ (where m$_p$ is bare proton mass).
Both assumptions, however, are not justified (see, e.g., TT95,
TT97). Also, their empirically constructed T$^{cr}_p$ is based on
the microphysical gap calculations carried out for `conventional'
neutron matter with only very small proton fraction. However, in
order for the nucleon direct URCA to work, the proton fraction
must be high, $\sim$ 15 \% or more (Lattimer et al. 1991). Then,
there will arise two important microphysical factors to {\it
suppress} the realization of proton superfluidity. One is that the
proton effective mass decreases with increasing $\rho$ and should
be less than $\sim 0.6 m_p$, even if such high proton
concentration phase could be realized at rather low densities
($\rho \gsim (2-3) \rho_0$). The other is that the repulsive core
effect in the $^1S_0$ pairing interaction, which grows with
increasing proton fraction, becomes more significant. The growth
of this effect means the reduction of attraction in the pairing
interaction, and hence a smaller energy gap.  These two factors
indeed {\it reduce} T$^{cr}_p$ to {\it below} 10$^8$ K for the
density regime where nucleon direct URCA can take place. We
emphasize that this qualitative tendency is {\it not}
model-dependent (TT97). Note that both effective mass and the
attractive effect in the pairing interaction depend sensitively on
composition, as well as density, -- these values for neutron
matter with low proton fraction (`conventional' neutron matter)
and high proton fraction (the matter which permits nucleon direct
URCA) should be quite different (TT97).

Furthermore, recent theoretical developments point to some
additional theoretical problems concerning the realization of the
nucleon direct URCA itself.  For instance, Engvik, et al. (1997)
(hereafter E97) and Akmal, Pandharipande and Ravenhall (1998)
(hereafter APR98) studied the symmetry energy in nuclear matter
adopting various new potentials (sometimes called the `modern
potentials'), and have shown that the direct URCA cannot occur, at
least below $\sim$ 5 times the nuclear density, for all
potentials.  (If the transition is to occur at such high density,
the EOS will have to be too soft, to be consistent with
observation.)  See also the very recent paper by Heiselberg \&
Pandharipande (2000) (hereafter HP00), which reviews the current
status of microphysical calculations of nuclear matter with the
modern potentials, and points out the consistency among various
results obtained by different methods, especially for neutron
matter. For instance, the results by E97 which is based on the
Brueckner theory, and those by APR98 which is based on a
variational calculation, are consistent.

The difficulties with the nucleon and kaon direct URCA options
just pointed out leave as still possible nonstandard options the
direct URCA involving pions, hyperons or quarks.  Hyperons may
appear at densities as low as $\sim(2-3)\rho_0$ (e.g., Prakash, et
al. 1992). Recently, Takatsuka and Tamagaki (1999, 2001) and
Takatsuka, et al. (2001) investigated hyperon superfluidity by
using several hyperon-hyperon pairing interactions, and conclude
that
hyperon cooling may be consistent with the observed detection data
of cooler stars.  Therefore we are currently exploring the hyperon
cooling option (Tsuruta et al. 2002).  The problem for quarks,
however, is that theoretically there are still too many unknown
factors to offer the level of exploration possible for the other
options (see Tsuruta, et al. 2002).

As to the pion direct URCA option, it has been already shown,
through detailed microphysical calculations, that T$^{cr}$ could
be several times 10$^9$ K, higher than the nucleon and kaon direct
URCA cases, and so quasi-baryons in the pion-condensed phase
should safely be in the superfluid state for the observed cooler
NSs. This is mainly because the effective mass of quasi-baryons is
higher in the pion-condensed phase, $\sim(0.8-0.9) m_N$, where
$m_N$ is the bare nucleon mass (TT82). Note that the superfluid
gap model adopted for the dashed curve in Fig. 1 is based on
realistic microphysical theories (TT82, U94, TT02). The conclusion
is that {\it the pion cooling is consistent with both theory and
observation.}

It may be noted that very recent theoretical developments indeed
assures the presence of pion condensates in the relatively low
density regime which is consistent with our pion cooling models.
For instance, Akmal \& Pandharipande (1997) carried out a
variational calculation of nuclear matter by using a modern
potential, and found that both symmetric nuclear energy matter and
pure neutron matter undergo transitions to pion condensed phase at
relatively low densities. Furthermore, adopting the most recent
experimental data on the giant Gamow-Teller resonance, Suzuki,
Sakai \& Tatsumi (1999) reexamined the threshold conditions for
pion condensation, and reached similar conclusion. See also the
review by HP00 which supports this conclusion.

\section{Summary and Concluding Remarks}

We have shown that the most up-to-date observed temperature data
are consistent with the current NS thermal evolution theories if
slightly less massive stars cool by standard cooling while
slightly more massive stars cool with nonstandard cooling.  Among
various nonstandard cooling scenarios, both nucleon and kaon
direct URCA cooling appears to be excluded due to the high proton
concentration required, which lowers the critical superfluid
temperature.  Then, the cooler stars must possess a central core
consisting of pions, hyperons or quarks. The pion cooling is
already shown to be consistent with both observation and theory.
Comparison of cooling curves with observations already eliminates
both very stiff and very soft EOS if the stellar mass is close to
1.4 M$_\odot$. That means the radius should be around 10 -- 12 km,
but not $\sim$ 7 km (soft EOS) nor $\sim$ 16 km (stiff EOS).

The capability of constraining the composition of NS interior
matter purely through observation alone will be limited, and hence
it will be very important to {\it exhaust all theoretical
resources.}  Theoretical uncertainties can be large, especially in
the supranuclear density regime, but here we emphasize that we
should still be able to set {\it acceptable ranges}, at least to
separate models more serious from those merely possible. More and
better data expected soon from {\it Chandra}, {\it XMM/Newton},
{\it HST}, and the 3rd generation missions already scheduled for
the near future, when combined with improved theories, should give
still better insight to some fundamental problems in dense matter
physics.

\acknowledgments

We thank Dr. Pavlov for helping us with obtaining the newest
observational data, and the anonymous referee for helpful
suggestions. ST's contribution was supported in part by NSF grant
PHY99-07949, NASA grant NAG5-3159, and a grant from the Yamada
Foundation.

\noindent
{\bf Figure Caption}

\noindent Fig.1.-- Cooling curves for stars with the medium FP
EOS. The surface photon luminosity which corresponds to surface
temperature (both to be observed at infinity) is shown as a
function of age. The solid curve shows standard cooling of a 1.2
M$_\odot$ NS with the T72 superfluid model, while the dashed curve
is the nonstandard pion cooling of a 1.4 M$_\odot$ star with the
E1-0.8 superfluid model. (See the text for the definitions and
further details.) The vertical bars refer to confirmed surface
temperature detection with error bars, for (1) RX J0822-4300, (2)
the Vela pulsar, (3) PSR 0656+14, (4) Geminga, (5) RX J1856-3754,
and (6) PSR 1055-52. The downward arrows refer to the temperature
upper limits for (a) Cas A point source, (b) Crab pulsar, (c) PSR
J0205+6449, (d) PSR 1509-58, (e) PSR 1706-44, (f) PSR 1823-13, (g)
PSR 2334+61, (h) PSR 1951+32, (i) PSR 0355+54, and (j) PSR
1929+10. The references are: Zavlin et al.1999 for (1), Pavlov et
al. 2001 for (2), Zavlin et al. 2002b for (3), Halpern \& Wang
1997 for (4), Zavlin et al. 2002a and Walter and Lattimer 2002 for
(5), Pavlov et al. 2002 for (6), Pavlov et al. 2000 for (a), and
Slane et al. 2002 for (c). The rest are taken from Becker 1995. In
spite of possible positive detections, sources 1E 1027.4-5209, RX
J0002+62, and RX J0720.4-3125 are not shown in Fig. 1 because
currently there are still some uncertainties including the age
estimate.

\clearpage

\begin{figure}
\plotone{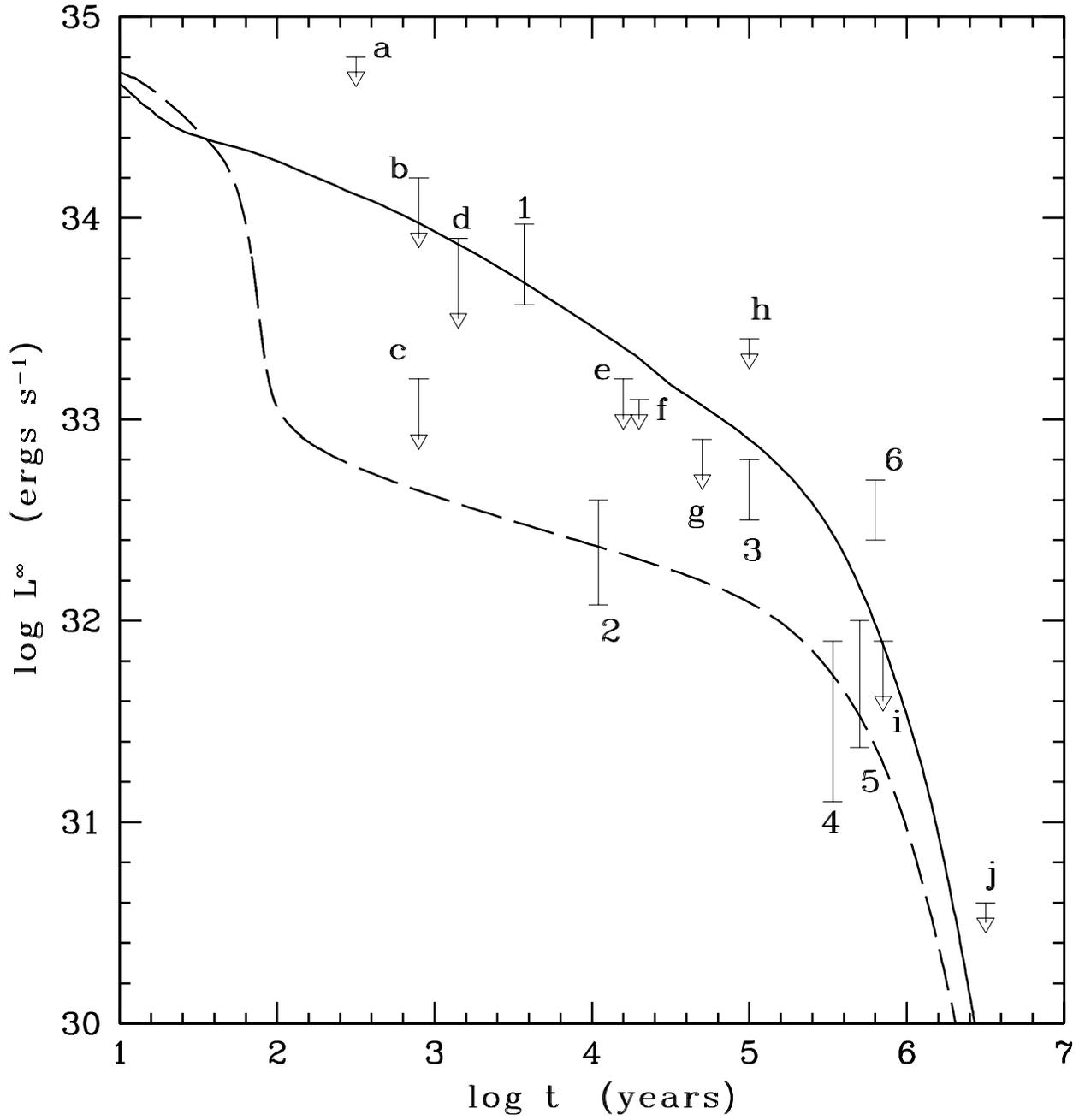}
\caption{see previous page \label{fig1}}
\end{figure}

\end{document}